\documentclass{elsart}
\usepackage{graphicx,epsfig}

\begin{document}
%%%%%%%%%%%%%%%%%%%%%%%%%%%%%%%%%%%%
\def\ltsimeq{\,\raise 0.3 ex\hbox{$ < $}\kern -0.75 em
 \lower 0.7 ex\hbox{$\sim$}\,}
 \def\gtsimeq{\,\raise 0.3 ex\hbox{$ > $}\kern -0.75 em
  \lower 0.7 ex\hbox{$\sim$}\,}

\def\avg #1{\langle #1\rangle}
%%%%%%%%%%%%%%%%%%%%%%%%%%%%%%%%%%%%
\begin{frontmatter}

\title{Pareto and Boltzmann-Gibbs behaviors in a deterministic 
multi-agent system}
\author{J. Gonz\'alez-Est\'evez$^{1,2}$, M. G. Cosenza$^2$, R. L\'opez-Ruiz$^3$, and J. R. S\'anchez$^4$}
\address{$^1$Laboratorio de F\'isica Aplicada y Computacional, 
Universidad Nacional Experimental del T\'achira, San Crist\'obal, Venezuela.\\
$^2$Centro de F\'isica Fundamental,  
Universidad de Los Andes, M\'erida, Venezuela.\\
$^3$DIIS and BIFI, Facultad de Ciencias, 
Universidad de Zaragoza, E-50009 Zaragoza, Spain.\\
$^4$Departamento de F\'isica, Facultad de Ingenier\'ia, 
Universidad Nacional de Mar del Plata, Mar del Plata 7600, Argentina.}

\date{\today}

\begin{abstract}
A deterministic system of interacting agents is considered as a model for economic dynamics. 
The dynamics of the system is described by a coupled map lattice with near neighbor interactions.
The evolution of each agent results from the competition between two factors: the agent's own 
tendency to grow and the environmental influence that moderates this growth. 
Depending on the values of the parameters that control these factors, the system can display Pareto 
or Boltzmann-Gibbs 
statistical behaviors in its asymptotic dynamical regime. 
The regions where these behaviors appear are calculated on the space of parameters
of the system. Other statistical properties, such as the mean wealth,
the standard deviation, and the Gini coefficient characterizing the degree of equity in the
wealth distribution are also calculated on the space of parameters of the system.
\end{abstract}
\end{frontmatter}

It is currently well established that income or wealth distribution in many western societies 
presents essentially two phases. This means that the society can be differentiated in 
two disjoint populations in which the probability distribution of wealth has a different functional form
in each of them \cite{yakovenko2001,Chatterjee05,Chakrabarti06,yakovenko2007,Chatterjee07}. Analysis of real 
economic data from U.K. and U.S.A. \cite{Dragulescu01} has shown that one phase possesses an exponential 
or Boltzmann-Gibbs probability distribution that involves about $95\%$ of individuals, mainly those with 
low and medium wealths,
and that the other phase, consisting of the the $5\%$ of individuals with highest wealths, shows a power law
distribution or Pareto behavior. Several economic models based on diverse probabilistic mechanisms for 
interaction between agents have been proposed in order to reproduce these types of statistical behavior 
\cite{yakovenko2000,lopezruiz2007,chakraborti2000,chakraborti2004,angle2006}. 
However, in most cases, both classes of distributions do not appear in a simple model; changes in the
interaction rules between agents are required in order to obtain either type of behavior. 

Randomness is an essential ingredient in all the former models.
Thus, agents behave as a classical gas without notion of locality \cite{Chatterjee07}. 
The interaction between agents 
occurs in pairs chosen at random, and these pairs exchange a random quantity of wealth in each 
transaction; this leads to an asymptotic state where the
wealth in the system follows a  Boltzmann-Gibbs distribution. The transition from a 
Boltzmann-Gibbs distribution to a Pareto behavior requires a change of structural properties
of the system. A power law distribution can be reached, for instance, by introducing a strong 
inhomogeneity in the properties of the agents \cite{yakovenko2007, chakraborti2004}. Thus, very different 
setups are needed in those models in order to simulate the collective behavior of real economic systems. 
On the other hand, interactions among real economic agents cannot be regarded as fully random. In fact, 
most economic transactions are driven by some kind of mutual interests or rational forces. 

In this article, we study the statistical properties of a recently introduced deterministic, 
network-based multi-agent dynamical model possessing minimal ingredients \cite{sanchez2007}. 
In particular, we  show that this simple model is capable of displaying Boltzmann-Gibbs as well as 
Pareto statistical behaviors in its asymptotic states.  

The system consists of $N$ agents placed at the nodes of a network. 
Each agent, representing an individual, a company, a country 
or other economic entity, is identified by an index $i$, with $i=1,\ldots,N$.  The dynamics of each 
agent is described by a discrete-time map that expresses the competition between its own tendency to 
grow and an environmental influence that controls
this growth. Although the model can be defined on any network of interacting agents, for simplicity 
we shall consider here a one-dimensional lattice with periodic boundary conditions.  
The dynamics of the system is described by the coupled map equations \cite{sanchez2007}
\begin{equation}
\begin{array}{ll}
x_{t+1}^i = &r_i\:x_t^i\: \exp(-\mid x_t^i-a_i\Psi_t^i\mid), \\
\Psi_t^i=&\frac{1}{2}(x^{i-1}_t+x^{i+1}_t),
\end{array}
\label{eq:system}
\end{equation}
where $x_t^i\geq 0$ gives the state of the agent $i$ at discrete time $t$, and it may denote the 
{\it wealth} of this agent; the factor  $r_ix_t^i$ expresses the {\it self-growth capacity} of agent $i$, 
characterized by a parameter $r_i$; $\Psi_t^i$ represents the local field acting at the site $i$ at 
time $t$; and $a_i$ measures the coupling of agent $i$ with its neighborhood; it can also be interpreted 
as the {\it local environmental pressure} exerted on agent $i$ \cite{ausloos2003}. The negative exponential 
function acts as a {\it control factor} that limits this growth with respect to the local field. 
With the dynamics given by Eqs.(\ref{eq:system}) the largest possibility of growth for agent $i$ is 
obtained when  $x_t^i\simeq a_i\Psi_t^i$, i.e., when the agent has reached some kind of adaptation 
to its local environment. 

\begin{figure}[tt]
\centerline{\includegraphics[width=8.6cm]{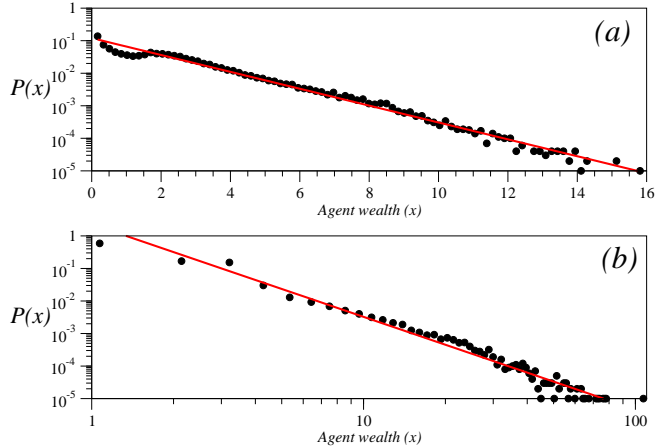}}
\caption{(a) Semilog plot of the probability distribution $P(x)$ of the states of the agents $x_t^i$
at time $t=10^4$, for $a=0.6$ and $r=4$. 
(b) Log-log plot of $P(x)$  at time $t=10^4$, for $a=0.92$ and $r=8$.} 
\label{fig1}
\end{figure}

For simplicity, in this paper we focus
on a homogeneous system where all agents possess the same growth capacity, $r_i=r$, and are subject to a uniform
selection pressure from their environment,  $a_i=a$. Thus, the parameter $a$ expresses the homogeneous wish of the
agents to reach a wealth level proportional to that of their environment. The value 
$a=1$ means a desire of being totally balanced with the neighborhood. The case
$a<1$ could be interpreted as some kind of lack of attitude in the population
for improving its relative wealth. When $a>1$ the agents possess
an excess of will (selfishness) for overcoming their local neighbors.

We study the collective behavior of the system described by Eqs.(\ref{eq:system}) in the space of parameters $(a,r)$.
For all the simulations shown, the system size is $N=10^5$ and the values of the initial conditions 
are uniformly distributed at random in the interval $x^i_0 \in [1,100]$. Also, a transient of $10^4$ iterations is 
discarded before arriving to the asymptotic regime where all the calculations are carried out. 
When indicated, time averages are done over the next $100$ iterations after the transient, and 
this result is newly averaged  over $100$ different realizations of the initial conditions with the same process.

Figure~1 shows the probability distribution of the states of the agents $P(x)$ at time $t=10^4$ for 
different values of the parameters $a$ and $r$. 
In Fig.~1(a), a semilog plot of $P(x)$ shows that, for the parameters used, the probability can
be well described by a Boltzmann-Gibbs distribution $P(x)\sim e^{-\mu x}$, where  $\mu=0.59$.
A thermodynamical simile can be established by defining
a kind of `temperature', $h=1/\mu=1.69$, that is related with the mean wealth of the agents
in the ensemble. For other values of the parameters,  $P(x)$ can display a Pareto-type behavior, 
as shown in the log-log plot of Fig.~1(b). In this case,  $P(x)\sim x^{-\alpha}$, with an exponent 
$\alpha=2.84$, a value in agreement with the exponents derived from real economic data 
\cite{yakovenko2001,levy1997,souma2001}. 

\begin{figure}[t]
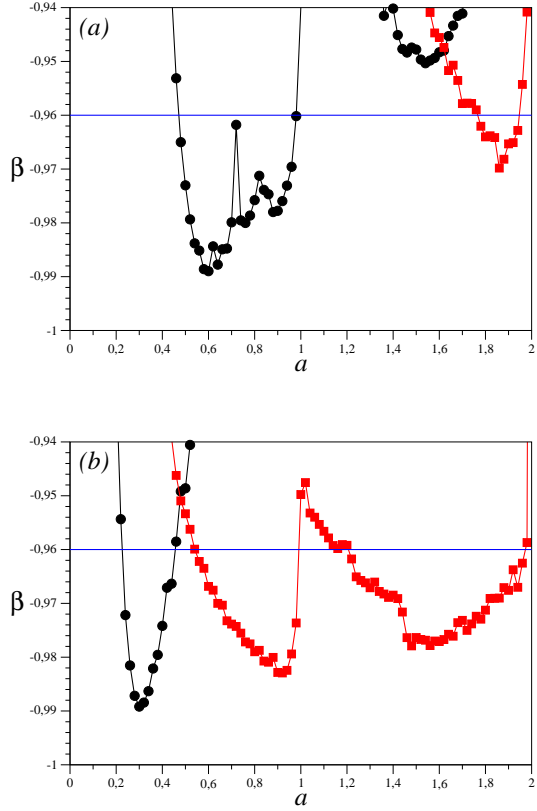

\centerline{\includegraphics[width=7cm]{fig2a.eps}}
\vspace{8mm}
\centerline{\includegraphics[width=7cm]{fig2b.eps}}
\caption{Correlation coefficient $\beta$ of the scaling exponents for the semilog fitting of the 
Boltzmann-Gibbs (circles) and for the log-log fitting of the
Pareto (squares) distributions, as a function of the parameter $a$ for two different values of $r$.  
The value $\beta=-0.96$ is indicated by a horizontal line. The values of $\beta$ shown are the result 
of averaging the values obtained over $100$ iterations for each parameter values, after discarding $10^4$ 
transients. (a) $r=4$. (b) $r=8$.}
\label{fig2}
\end{figure}

The exponents $\mu$ and $\alpha$ in the distributions shown in Fig.~1 are obtained by linear regression using 
the least-squares method; this gives a value of the correlation coefficient greater than $0.98$ in each case. 
Boltzmann-Gibbs and Pareto distributions also appear for other values of the parameters $(a,r)$.  
Figure~2 shows the correlation coefficient $\beta$ corresponding to the semilog fitting of the Boltzmann-Gibbs 
as well as the log-log fitting of the Pareto distributions as a function of the parameter $a$, 
for two different values of $r$. We consider that
either of these fittings are accurate enough when $|\beta|>0.96$. The intervals of the parameter $a$ 
where this condition has been achieved can be identified in Fig.~2.

Figure~3 shows the regions where the probability distribution $P(x)$ displays Boltzmann-Gibbs and Pareto 
behaviors in the space of parameters $(a,r)$ for the system Eqs.(\ref{eq:system}). Both parameters $a$ and $r$ 
are varied in intervals of size $0.02$ and for each pair $(a,r)$ the semilog and the log-log linear 
regressions described in Fig.~1 are performed after discarding $10^4$ transients and averaging over the 
following $100$ iterations, and only those results yielding a correlation coefficient $|\beta|>0.96$ 
are shown in Fig.~3.

\begin{figure}[h]
\begin{center}
\includegraphics[width=7cm]{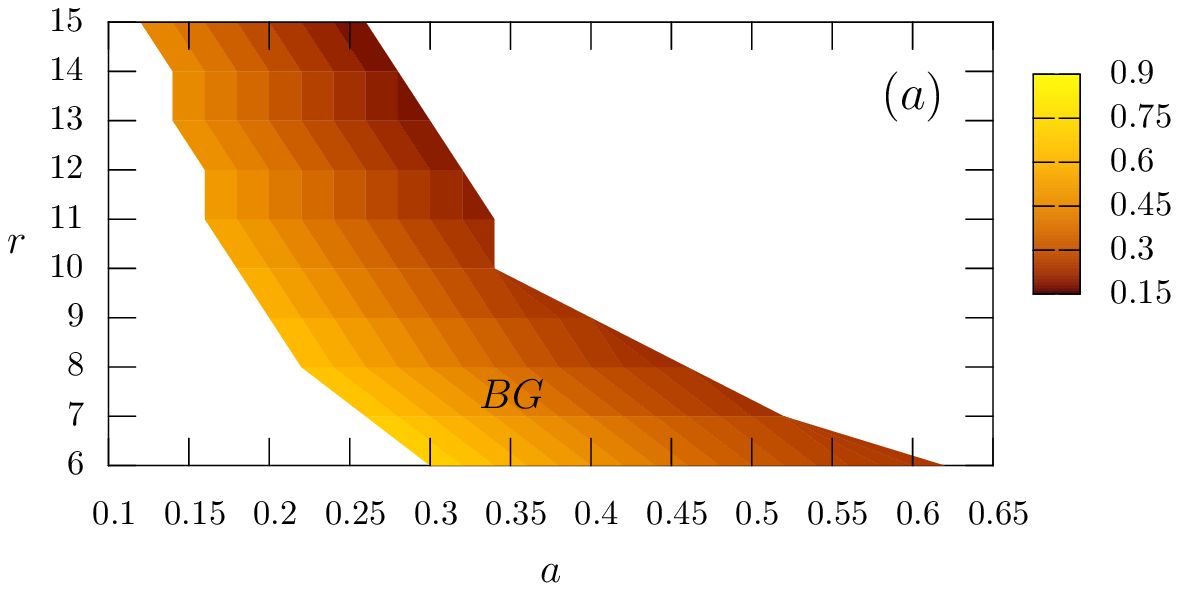}
\end{center}
\begin{center}
\includegraphics[width=7cm]{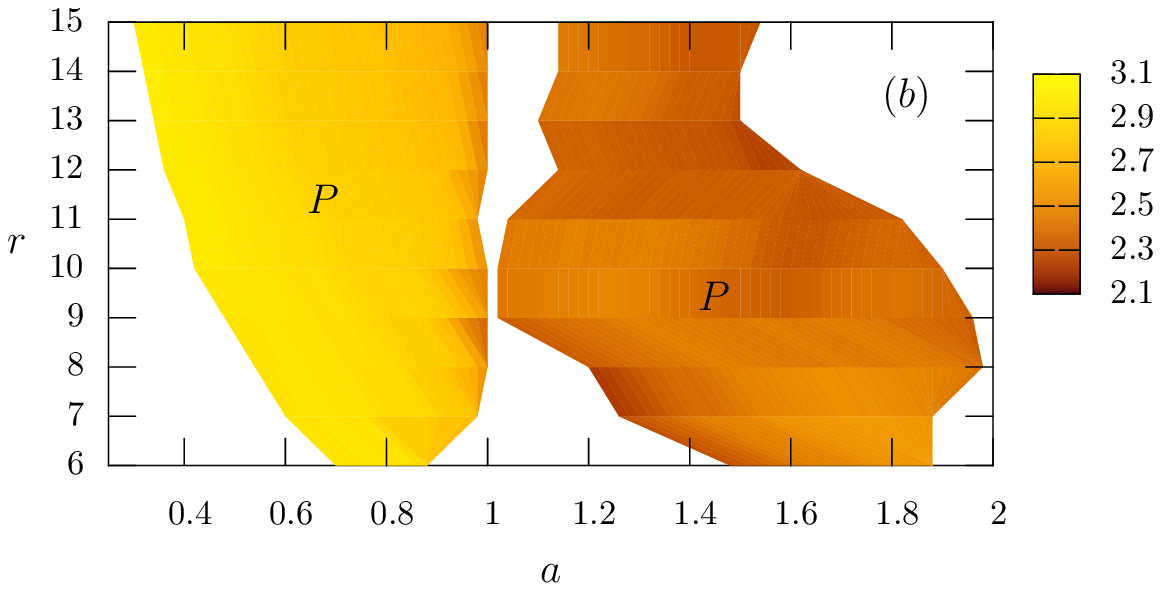}
\end{center}
\caption{(a) Regions where Boltzmann-Gibbs behavior $P(x)\sim e^{-\mu x}$ appears on the space of parameters 
$(a,r)$ are indicated by the label \textit{BG}. The color code on the right indicates the values of the scaling 
exponent $\mu$ obtained by the procedure explained in Fig.~1(a) and fullfilling the condition $|\beta|>0.96$. 
(b) Regions where Pareto behavior $P(x)\sim x^{-\alpha}$ occurs on the plane $(a,r)$ are labeled by \textit{P}.  The 
color code on the right indicates the values of the exponent $\alpha$ obtained as in Fig.~1(b) and 
satisfying $|\beta|>0.96$. In both cases, for each pair of values $(a,r)$ calculations of the scaling exponents 
are performed after discarding $10^4$ transients and averaging over the following $100$ iterations.}
\label{fig3}
\end{figure}

We note that Boltzmann-Gibbs behavior is found for 
lower values of the local environmental pressure $a$.  When the value of $a$ increases,
the population enters in a competitive regime that provokes the
appearance of the Pareto behavior in the system. The scaling exponents obtained for the Pareto behavior observed
in Fig.~3(b) are in the range $\alpha\in [2.3,3.0]$; these values are similar to those
found in actual economic data \cite{yakovenko2001,levy1997,souma2001}.

\begin{figure}[h]
\centerline{\includegraphics[width=7cm]{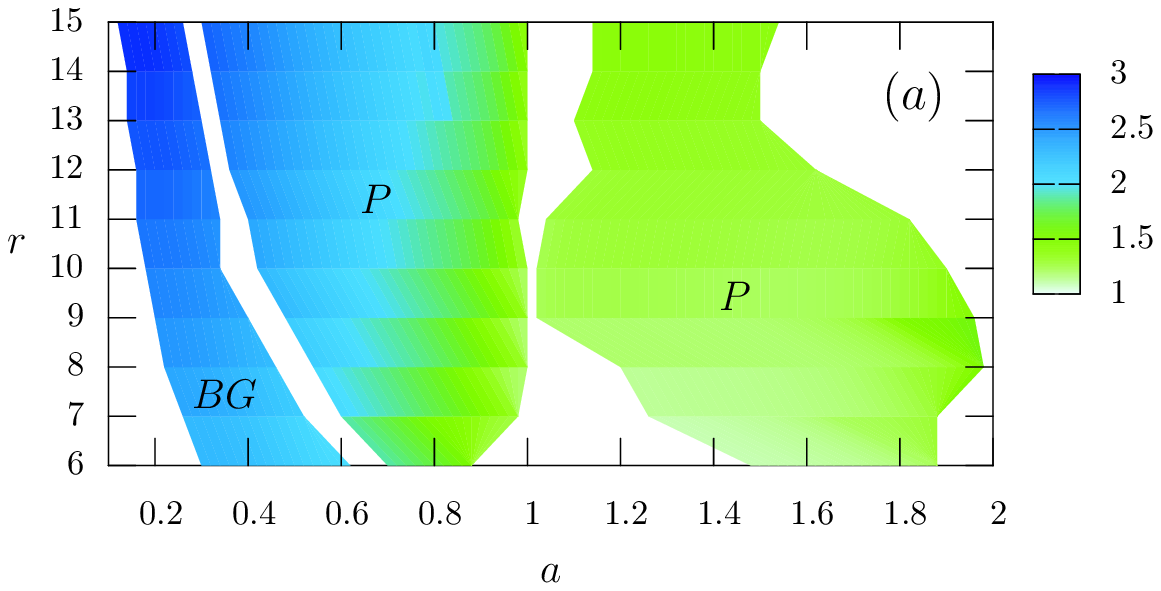}}
\centerline{\includegraphics[width=7cm]{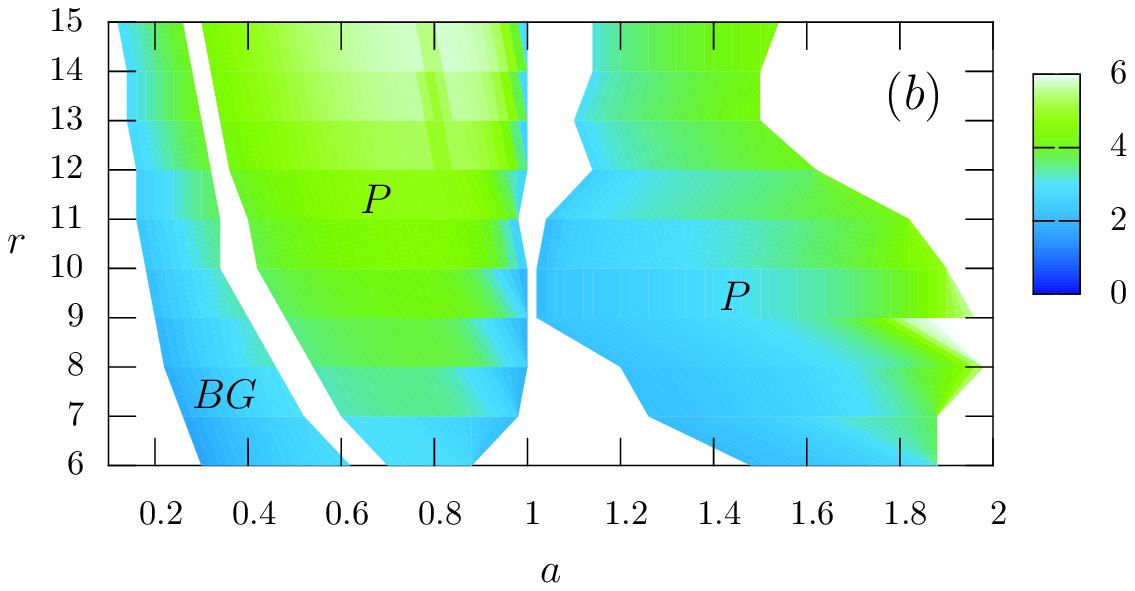}}
\caption{(a) Mean field $H_t$ of the system at $t=10^4$ for the regions where Boltzmann-Gibbs and Pareto 
behaviors are found on the plane $(a,r)$, indicated respectively by the labels \textit{BG} and \textit{P}. 
The color code on the right indicates the values taken by $H_t$. (b) The quantity $\langle\sigma\rangle$ 
on the space of parameters $(a,r)$. 
The color code on the right indicates the values of $\langle\sigma\rangle$.}
\label{fig4}
\end{figure}

The mean field of the system or average wealth per agent at a time $t$ is defined as 
\begin{equation}
H_{t}=\frac{1}{N}\sum_{i}^{N} x_t^i,
\end{equation} 

Figure~4(a) shows the asymptotic value of the mean field for the regions where Boltzmann-Gibbs and Pareto behaviors 
are observed on the space of parameters $(a,r)$.
Note that, although the values of the initial states of the agents are randomly distributed on the interval
$[1,100]$, the system evolves to an asymptotic state where $H_t$ takes values
on the smaller interval $[0,3]$. On the other hand, for some values of the parameters $a$ and $r$, the states of 
agents in the system at a given time exhibit a large dispersion. Similarly, for those parameters, the values of
the state of any agent present large fluctuations over long times. To characterize these fluctuations, we define 
the instantaneous standard deviation of the mean field as
\begin{equation}
\sigma_{t}=\left( \frac{1}{N}\sum_{i=1}^N\left[x_t^i-H_t \right]^{2} \right)^{1/2}.
\end{equation} 
After discarding $10^4$ transients, we calculate the mean value 
of $\sigma_t$  over $100$ iterations, and then average this result over $100$ realizations of initial conditions. 
The resulting average dispersion, denoted by $\langle  \sigma \rangle$, is shown in Fig.~4(b) on the plane $(a,r)$ 
for the same regions indicated in Fig~4(a). Note that for some regions of parameters the quantity 
$\left\langle\sigma_t\right\rangle$ can be much greater than the value of the mean field.
For instance, about the line $a=1.9$, the mean field
is small,  $H_t \approx 1$, but $\langle\sigma_t\rangle \approx 6$, showing that the fluctuations can be very large
in this system. Thus, in spite of its simplicity, the deterministic model given by Eqs.~(\ref{eq:system}) 
can exhibit great spatiotemporal complexity. 

\begin{figure}[h]
\centerline{\includegraphics[width=8cm]{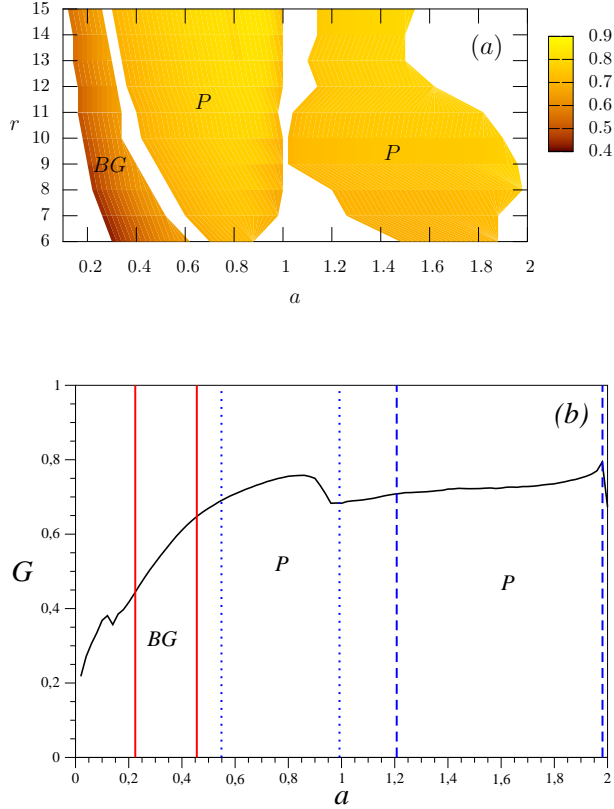}}
\vspace{10mm}
\centerline{\includegraphics[width=8cm]{fig5b.eps}}
\caption{(a) Gini coefficient at $t=10^4$ as a function of the parameters $a$ and $r$.
The labels \textit{BG} and \textit{P} indicate the regions where Boltzmann-Gibbs and Pareto behaviors are 
observed on the plane $(a,r)$.
The color code on the right indicates the values taken by the Gini coefficient.
(b) Gini coefficient vs. $a$, with fixed $r=8$, where the labels \textit{BG} and \textit{P} correspond to 
the regions indicated in (a).}
\label{fig5}
\end{figure}

The large dispersions observed in Fig.~4(b) reflect the inequity in the wealth distribution among agents in 
the system. To characterize the degree of inequality in the wealth distribution we use the Gini coefficient 
defined at a time $t$ as \cite{gini}
\begin{equation}
G_t=\frac{1}{2N^2 H_t} \sum_{i,j=1}^N |x_t^i - x_t^j|.
\end{equation}
A perfectly equitable  distribution of wealth at time $t$, where $x_t^i=x_t^j, \forall i,j$, yields a value 
$G_t=0$. The opposite situation, where one agent has the total wealth $\sum_{i=1}^N x_t^i$, has a value of 
$G_t=1$. Figure~5(a) shows
the asymptotic value of the Gini coefficient on the plane of parameters $(a,r)$. 
Note that the Gini coefficient reaches larger values, i.e.  $G_t \in [0.6,0.8]$, in the regions associated to 
Pareto regimes, while it takes lower values, i.e. $G_t \in[0.4,0.6]$, in the region corresponding to 
Boltzmann-Gibbs behavior. This results agree with our qualitative understanding that equity is more favored 
in the presence of a larger middle economic class in a society, as expressed by a Boltzmann-Gibbs distribution. 
A plot of $G_t$ as a function of $a$ for a fixed value $r=8$ is shown in Fig.~5(b), where the Boltzmann-Gibbs 
and Pareto regions are also indicated.

In summary, the deterministic model Eqs.~(\ref{eq:system})
shows statistical behaviors described by
Boltzmann-Gibbs and Pareto distributions in different regions of its parameters.
The appearance of these collective properties does not require the addition of ramdonness or
any structural change in the system. Only some appropriate tuning of the parameters of the system is needed
to obtain either type of behavior. This property contrasts with most models for economic behavior in the literature 
which require changes in their dynamical rules in order to yield an exponential or a power law distribution
of states. Since it is currently accepted that most western societies consist of two
differentiated economic classes characterized by different distribution functions \cite{yakovenko2007}, 
coupled map models such as Eqs.~(\ref{eq:system}) can be useful to study the formation of these two economic 
populations. Our results support the view that determinism alone can give rise to some relevant collective 
behaviors observed in economic systems. This basic model can be readily extended to include the considerations 
of more complex networks of interactions, heterogeneities, and the coevolution of dynamics and the topology of 
connectivity, among other interesting issues.

\section*{Acknowledgments}
This work was supported in part by Decanato de Investigaci\'on of the Universidad Nacional Experimental 
del T\'achira (UNET), under grants 04-001-2006 and 04-002-2006. J.G.-E. thanks Decanato de Investigaci\'on 
and Vicerrectorado Acad\'emico of UNET for travel support.  J.G-E. and R.L-R. acknowledge support from BIFI, 
Universidad de Zaragoza, from Asociaci\'on Iberoamericana de Postgrado (AUIP), and by grant 
DGICYT-FIS2006-12781-C02-01, Spain. 
M. G. C. is supported by grant C-1396-06-05-B  from Consejo de Desarrollo, 
Cient\'ifico, Tecnol\'ogico y Human\'istico, Universidad de Los Andes, M\'erida, Venezuela.

\end{document}